\documentclass[aps,prl,twocolumn,superscriptaddress,groupedaddress]{revtex4} 
\usepackage{graphicx}  % needed for figures
\usepackage{dcolumn}   % needed for some tables
\usepackage{bm}        % for math
\usepackage{amssymb}   % for math

\usepackage{amsmath}
\usepackage{dsfont}
\usepackage{caption}
\usepackage{xcolor}

\newcommand{\grafe}[1]{\left\{ #1 \right\}}
\newcommand{\tonde}[1]{\left( #1 \right)}
\newcommand{\quadre}[1]{\left[ #1 \right]}

\begin{document}

\title{Remanent magnetization: signature of Many-Body Localization in quantum antiferromagnets}

\author{V. Ros}
\email{vros@sissa.it}
\affiliation{SISSA- International School for Advanced Studies, via Bonomea 265, 34136 Trieste, Italy}
\affiliation{INFN Sezione di Trieste, Via Valerio 2, 34127 Trieste, Italy}

\author{M. M{\"u}ller}
\email{Markus.Mueller@psi.ch}
\affiliation{Paul Scherrer Institut, CH-5232 Villigen PSI, Switzerland}
\affiliation{The Abdus Salam International Center for Theoretical Physics, Strada Costiera 11, 34151 Trieste, Italy}
\affiliation{Department of Physics, University of Basel, Klingelbergstrasse 82, CH-4056 Basel, Switzerland}

%\date{\today}

\begin{abstract}
We study the remanent magnetization in antiferromagnetic, many-body localized quantum spin chains, initialized in a fully magnetized state. Its long time limit is an order parameter for the localization transition, which is readily accessible by standard experimental probes in magnets. We analytically calculate its value in the strong-disorder regime exploiting the explicit construction of quasi-local conserved quantities of the localized phase. We discuss analogies in cold atomic systems.
\end{abstract}

\maketitle

\paragraph{Introduction.}
The non-equilibrium dynamics in disordered, isolated quantum systems have been subject to theoretical investigations ever since the notion of localization was  introduced in~\cite{anderson1958absence}. Spin systems in random fields are prototypical models to analyze the disorder-induced breakdown of thermalization: a large number of studies on disordered spin chains~\cite{oganesyan2007localization, vznidarivc2008many, Bauer:2013rw, de2013ergodicity, mondragon2015,pal2010MBL,Kjall:2014fj,Vosk:2013kq, alet2015, John2015TotalCorrelations, Luitz2016LongTail, Luitz2016ExtendedSlow} has provided evidence for a dynamical phase transition between a weak-disorder phase which thermalizes, and a Many-Body Localized (MBL) phase in which excitations do not diffuse, ergodicity is broken and local memory of the initial conditions persists for infinite time~\cite{AndersonFleishman82, gornyi2005interacting, Basko:2006hh}.

Signatures of MBL are found in the properties of individual many-body eigenstates. Even highly excited eigenstates exhibit area-law scaling of the bipartite entanglement entropy~\cite{Bauer:2013rw, pal2010MBL, Kjall:2014fj, Friesdorf2015} and Poissonian level statistics~\cite{oganesyan2007localization, Avishai2002LevelStatistics, Modak2014LevelStat}, both being incompatible with thermalization~\cite{Deutsch1991, Srednicki1994, RigolOlshanii}. Novel dynamical properties such as the logarithmic spreading of entanglement have been observed in direct simulations of the time evolution %\VR{cite the older paper?}
\cite{vznidarivc2008many,bardarson2012unbounded}. The non-equilibrium physics of MBL systems has  been probed experimentally in artificial quantum systems made of cold atomic gases~\cite{Bloch2015, schneider2015} and trapped ion systems~\cite{monroe2015}, while an indirect signature in the from of strongly suppressed absorption of radiation was found in electron-glasses~\cite{Ovadyahu2012suppression}. However, direct observations of MBL in solid-state materials are still lacking.

It has been argued~\cite{huse2013phenomenology,serbyn2013local,rms-IOM} that the properties of MBL systems are related to the existence of extensively many quasi-local conserved operators that strongly constrain the quantum dynamics, preventing both transport and thermalization. Their existence also follows as a corollary from Imbrie's rigorous arguments in favor of MBL~\cite{imbrie2014many,OganesyanHuseIntegrals}.

\begin{figure}[htbp]
  \centering
  \includegraphics[width=1\linewidth]{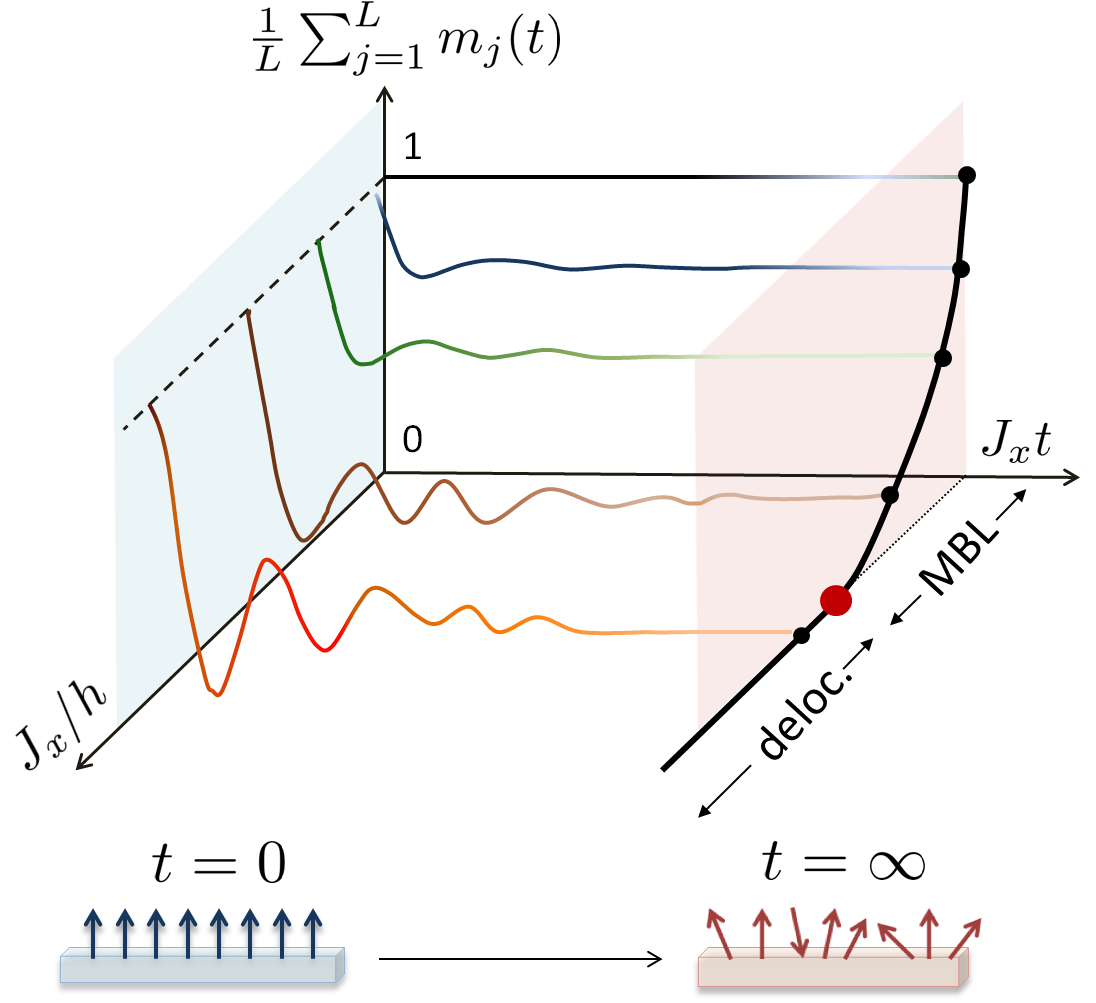}
  \captionof{figure}{\small{Relaxation of the total magnetization from a fully polarized initial state. The black curve is the stationary value $L^{-1}\sum_{j} \hat{m}_j$: it vanishes at the critical point separating the MBL and delocalized phases (red point), and it is non-analytic for $J_x/h \ll 1$, cf. Eq.~\eqref{RatioSF}.}}
  \label{fig:pictorial}
\end{figure}

In this work, we propose a experimentally readily observable consequence of MBL in quantum magnets: the out-of-equilibrium remanent magnetization that persists after ferromagnetically polarizing an antiferromagnet whose  total magnetization is not a conserved. The remanence implies non-ergodicity, since ergodic dynamics would relax the magnetization completely (cf. Fig.~\ref{fig:pictorial} for a schematic sketch of the protocol). As an example, we consider an antiferromagnetic, anisotropic Heisenberg spin-$1/2$ chain
\begin{equation}\label{SpinChainInt}
 H=\sum_{k} \tonde{h_k \sigma_k^z - \sum_{\alpha= x,y,z}J_\alpha \sigma_k^\alpha \sigma_{k+1}^\alpha}
\end{equation}
subject to random fields $h_k$ along the Ising axis. We assume $J_z<0$, as well as $J_x\neq J_y$ to ensure the non-conservation of the total magnetization. Such Hamiltonians can be realized, e.g., in Ising compounds with both exchange and dipolar interactions.  However, essentially any sufficiently strongly disordered quantum antiferromagnet with non-conserved magnetization should exhibit qualitatively the same phenomenology as the chains described here.
 
 The remanent magnetization serves as an order parameter for the dynamical phase transition. It is a magnetic analogue of the remanent density modulation considered in \cite{Andraschko2014purification} and measured in recent cold-atom experiments~\cite{Bloch2015, schneider2015}. It should be experimentally much simpler to access since it focuses on the total magnetization (at $q=0$) which can be readily picked up by a squid, without requiring scattering measurements to resolve spatial patterns. 
 
% For strong disorder, we explicitly construct the local integrals of motion (dressed spin operators), as  developed in~\cite{rms-IOM}, and use them to calculate analytically the remanence.  

\paragraph{Conserved dressed spins.} We consider random fields $h_k$ uniformly distributed in $\quadre{-h, h}$, and assume strong anisotropy of the couplings, $|J_y| \ll |J_x| \ll |J_z|, h $. For simplicity we restrict to $J_y=0$.
For $J_x=0$, the spin chain is classical and trivially localizes dynamically, as the $\sigma^z_k$ form a complete set of commuting, local, conserved operators. The eigenstates are product states in this basis. For sufficiently small $|J_x| \ll h$ localization is predicted to remain intact, coming along with a complete set of conserved and mutually commuting, but now dressed spin operators $I_k= \sigma^z_k + O(J_x/h)$, or ``l-bits".

Unlike the conserved charges of one-dimensional integrable systems, the operators $I_k$ are quasi-local: their action decays exponentially away from the localization center $k$, $|| [I_k , \sigma_j^\alpha]|| \leq {\rm const}\times \exp[- |j-k|/\xi]$ with a finite correlation length $\xi$. 
Below, following Ref.~\cite{rms-IOM}, we explicitly construct these dressed spin operators to low orders in $J_x$, and use them to calculate analytically the remanent magnetization.  

We assume that by applying a strong field, the antiferromagnetic chain was prepared in the fully magnetized state $|\psi_0\rangle$ with density matrix:  \begin{equation}\label{neel}
 |\psi_0\rangle \langle \psi_0|= \prod_{i} \frac{1+ \sigma_i^z}{2}.
\end{equation}
After switching off the field the dynamics is governed by~\eqref{SpinChainInt}. \footnote{One may also view this protocol as a quantum quench: First, a high energy eigenstate of the Hamiltonian with $J_x=0$ is prepared and then the quantum fluctuations $J_x \sigma^x_k \sigma^x_{k+1}$ are switched on abruptly at time $t=0$.}

We are interested in the long time behavior of the magnetization, and thus consider the time averaged magnetization at site $j$:
\begin{equation}\label{LocalMag}
% \hat{m}_j(\omega)= \lim_{T\to \infty} \frac{1}{T}\int_0^T \frac{dt}{2 \pi} e^{-i \omega t} \langle \psi_0 | \sigma^z_j (t)| \psi_0 \rangle.
 \hat{m}_j= \lim_{T\to \infty} \frac{1}{T}\int_0^T dt  \,m_j(t); \,\,
 %= \lim_{T\to \infty} \frac{1}{T}\int_0^T dt  
 m_j(t) = \langle \psi_0 | \sigma^z_j (t)| \psi_0 \rangle.
\end{equation}

For $J_x=0$, the local magnetization is trivially conserved, ${m}_j(t)=1$. For finite $J_x$, $\sigma^z_j(t)$ has a non-trivial time dependence, which reduces $\hat{m}_j$. In the MBL regime, however, the time evolution is strongly constrained by the conservation of dressed spins $I_k$ with $|k-j|\lesssim \xi$. As a consequence, partial memory of the initial order $\langle \sigma^z_j \rangle=1$ is retained for arbitrarily long time, resulting in a finite remanence of the site-averaged magnetization $\hat{m} = L^{-1}\sum_j \hat{m}_j$.

In the absence of spectral degeneracies,~\eqref{LocalMag} can be expressed via a Lehmann representation as
\begin{equation}\label{def3y}
\hat{m}_j=  \sum_{\alpha} \langle \psi_0| P_\alpha \sigma^z_{j} P_\alpha| \psi_0 \rangle,
\end{equation}
where 
%\begin{equation}\label{projEigen}
$P_\alpha=| \psi_\alpha \rangle \langle \psi_\alpha |= \prod_{k=1}^L \tonde{1+i_k^{(\alpha)} I_k}/2$
%\end{equation}
projects onto the eigenstate labeled by the quantum numbers $i_k^{(\alpha)} \in \grafe{\pm 1}$ of the dressed spins $I_k$. Using the operator identity
\begin{eqnarray}\label{CommutatorsFormula}
   &&\sum_\alpha  P_\alpha \sigma^z_{j} P_\alpha= \sigma^z_j+ \\
   &&\sum_{n=1}^{L} \sum_{k_n>k_{n-1} \cdots >k_1} \prod_{l=1}^n \tonde{\frac{I_{k_l}}{2}}\quadre{ \quadre{\quadre{\sigma^z_j, I_{k_1}}, I_{k_2}}, \cdots, I_{k_n}},\nonumber
\end{eqnarray}
we obtain (cf. Appendix A for details):
\begin{widetext}
\begin{equation}\label{finalExp}
%\begin{split}
 \hat{m}_j= 1+ \sum_{n=1}^{L} \, \sum_{k_n>k_{n-1} \cdots >k_1} \text{Tr} \grafe{
 \prod_{i=1}^n \tonde{\frac{I_{k_i}}{2}}\quadre{ \quadre{\quadre{\sigma^z_j, I_{k_1}}, I_{k_2}}, \cdots, I_{k_n}} \prod_{i=1}^L \tonde{\frac{1 +\sigma^z_i}{2}}},
% \end{split}
\end{equation}

\end{widetext}
where $\text{Tr} \grafe{\cdot}$ denotes the trace, and an ordering among the labels of the operators $I_k$ is assumed \footnote{In the perturbative setting, there is a natural mapping between the set of conserved operators $I_k$ and the sites $k$, since $I_k$ is a perturbation of $\sigma^z_k$.}.

%\paragraph{Perturbative calculation.}
The expression~\eqref{finalExp} is particularly suitable for perturbative calculations. At any order, the terms $\delta I^{(n)}_k=O(J_x^n)$ in the asymptotic series:
\begin{equation}\label{formalPT}
 I_k = \sigma_k^z + \delta I^{(1)}_k+ \delta I^{(2)}_k + \cdots \,
\end{equation}
are uniquely fixed, by the constraints $\quadre{I_k, H}=0$ and $I^2_k=1$~\cite{rms-IOM}, see also Appendix B. For the Hamiltonian~\eqref{SpinChainInt} with $J_y=0$, the first order terms read
\begin{equation}\label{op1}
\begin{split}
\delta I^{(1)}_k &= \sum_{\rho, \tau= \pm 1} \tonde{A^{(k)}_{\rho \tau} O^{(k)}_{\rho \tau}-A^{(k-1)}_{\rho \tau} O^{(k-1)}_{\rho \tau}}\\
 &+\sum_{\rho, \tau= \pm 1}\tonde{B^{(k)}_{\rho \tau} \Delta^{(k)}_{\rho \tau}+B^{(k-1)}_{\rho \tau} \Delta^{(k-1)}_{\rho \tau}},
 \end{split}
\end{equation}
where we define the local operators
\begin{equation}\label{op2}
\begin{split}
 O^{(k)}_{\rho \tau}&= \frac{1 + \rho \, \sigma^z_{k-1}}{2}\quadre{\sigma^+_k \sigma^-_{k+1}+ \mathrm{h.c.}}\frac{1 + \tau \, \sigma^z_{k+2}}{2},\\
   \Delta^{(k)}_{\rho \tau}&= \frac{1 + \rho \, \sigma^z_{k-1}}{2}\quadre{\sigma^+_k \sigma^+_{k+1}+ \mathrm{h.c.}}\frac{1 + \tau \, \sigma^z_{k+2}}{2},
 \end{split}
\end{equation}
and the coefficients
\begin{equation}\label{op3}
 \begin{split}
  A^{(k)}_{\rho \tau}&=-\frac{J_x}{h_{k}-h_{k+1}+ J_z(\tau-\rho) },\\
    B^{(k)}_{\rho \tau}&=-\frac{J_x}{h_{k}+h_{k+1}- J_z(\tau+ \rho) }. 
\end{split}
\end{equation}

At low orders, the sum over multi-indices in Eq.~\eqref{finalExp} reduces to the few terms involving indices sufficiently close to $k$, since other commutators vanish. The lowest order corrections to $\hat{m}_j$ are given by the terms with $n=1,2$ in~\eqref{finalExp}. 
Inserting~\eqref{op1} into (\ref{CommutatorsFormula}) and (\ref{def3y}) we find \footnote{The amplitudes $A^{(j)}_{\rho \tau}$ do not contribute at this order due to the particular choice of the initial state.}:
\begin{equation}\label{def4y}
\begin{split}
 \hat{m}_j=1-\tonde{B^{(j)}_{1 ,1}}^2 -\tonde{B^{(j-1)}_{1, 1}}^2 + O(J_x^3). 
 \end{split}
 \end{equation}

The average $\langle \hat{m}_j \rangle_{\rm dis}$ over random fields (or, equivalently, over sites) is an analytic function of the couplings for $|J_z|>h$, while it is ill-defined for $|J_z|<h$; the apparent divergence is due to rare realizations of local fields that give rise to arbitrarily small energy denominators in Eq.~\eqref{op3}. Those occur when there are nearly degenerate (resonant) classical configurations that strongly hybridize via the exchange interaction $ J_x \sigma^x_k \sigma^x_{k+1}$. The resonant many-body configurations are also responsible for the divergence of the formal expansion~\eqref{formalPT} \footnote{One can verify that a small denominator generated at a given order in the series expansion for $I_k$ necessarily re-appears repeatedly in higher order terms, giving rise to subsequences of operators that are divergent in norm.}. In the MBL phase, however, the probability of resonant hybridization decays sufficiently fast with the distance between hybridizing degrees of freedom~\cite{imbrie2014many}. The divergent subsequences in~\eqref{formalPT} can then be re-summed~\cite{anderson1958absence}, yielding a `renormalized', norm-convergent operator expansion. 
 
The leading resonances can be re-summed by considering the simpler Hamiltonian 
\begin{equation}\label{Hreduced}
\begin{split}
   H^{(k)} &\equiv \sum_{i=1}^L \tonde{h_i \sigma_i^z - J_z \sigma_i^z \sigma_{i+1}^z}-J_x \sigma^x_k \sigma^x_{k+1},
 \end{split}
\end{equation}
where only one (resonant) $J_x-$coupling is retained. For this case a full set of exactly conserved operators $\tilde{I}_i$ satisfying $\tilde{I}_i^2= \mathds{1}$ can be constructed explicitly. It amounts to finding a local rotation that maps the $\sigma^z_k, \sigma^z_{k+1}$ to two operators $\tilde{I}_k, \tilde{I}_{k+1}$, and thus resums all perturbative terms containing higher powers of the resonant $J_x-$coupling. The two resulting "l-bits" contain the terms (\ref{formalPT}-\ref{op2}) (without higher order corrections), albeit with modified coefficients~\eqref{op3} given in Appendix B.
Together with the $ \tilde{I}_i=\sigma_i^z$ for $i \neq k, k+1$ they serve as a new basis for the perturbation theory in the remaining, non-resonant $J_x-$couplings.

Inserting these integrals into (\ref{def3y}), we find again an expression like (\ref{def4y}), but with the substitution:
\begin{equation}
  B^{(j)}_{\rho \tau} \longrightarrow -\frac{J_x}{\tonde{\quadre{h_{j}+h_{j+1}- J_z(\tau+\rho )}^2 + J_x^2}^{1/2}}.
\end{equation}
From this we obtain the remanent magnetization 
\begin{equation}\label{RatioSF}
\langle \hat{m}_j\rangle_{\text{dis}}=1- \frac{\pi |J_x|}{h} \tonde{1+\frac{J_z}{h}}+ O(J_x^2),
\end{equation}
which  for $|J_z|<h$ is non-analytic in $J_x$. This feature arises due to resonances. The non-analytic cusp at $J_x=0$ has the largest magnitude in the limit of vanishing Ising interactions, $J_z \to 0$ (recall that $J_z<0$), which corresponds to lowest effective disorder.

\paragraph{Atomic analogues.}
The gauge transformation 
$\mathcal{U}= \prod_{j=1}^L \text{exp}\tonde{i \frac{\pi}{2} j \sigma^x_j}$ maps the antiferromagnetic chain~\eqref{SpinChainInt} into its ferromagnetic counterpart with $J_x \to J_x$, $J_{y,z} \to -J_{y,z}$, and the initial state $|\psi_0 \rangle$ into a N\'{e}el state. The order parameter is mapped into the staggered magnetization. Such a quantity has been studied numerically in~\cite{DasSarma2016} for disordered, long-range transverse field Ising chains, modeling the ion-trap quantum simulators explored experimentally in~\cite{monroe2015}. The staggered magnetization is a close analogue of the particle imbalance studied as an experimental probe of many-body localization in cold atoms~\cite{Bloch2015}:
\begin{equation}\label{ImbalanceFermions}
 \mathcal{I}(t)= \frac{2}{L} \sum_{j=1}^L (-1)^j \langle n_j(t) \rangle.
\end{equation} 
Here $n_j$ is the occupation number of site $j$, which are prepared in an initial density wave $n_j(t=0)= [1+(-1)^j]/2$.

A ferromagnetic spin chain with $J_x= J_y$ is equivalent, via the Jordan-Wigner transformation, to a one-dimensional model of interacting spin-less fermions in a disordered potential. For $J_z=0$ it reduces to the non-interacting Anderson model 
\begin{equation}\label{Anderson}
H= - J \sum_{i=1}^{L-1} \tonde{c^\dag_i c_{i+1}+ h.c.} + 2 \sum_{i=1}^L h_i n_i
\end{equation}
for which the imbalance is a sum over single particle contributions, weighted with the occupation probability of eigenstates in the initial state. A standard calculation leads to the remanent imbalance~\eqref{ImbalanceFermions} in the form
\begin{equation}\label{ImbalanceAnderson}
  \hat{\mathcal{I}}=  \frac{1}{L} \sum_{\alpha=1}^L \tonde{\sum_{k=1}^L (-1)^{k} \phi^2_\alpha (k)}^2, 
\end{equation}
where $\phi_\alpha(i)$, with $1\leq \alpha, i \leq  L$ are the localized single particle eigenstates of the quadratic Hamiltonian~\eqref{Anderson}. This solvable case is interesting as it can be analyzed deeper into the weak disorder limit.

\begin{figure}[htbp!]
  \centering
  \includegraphics[width=1\columnwidth]{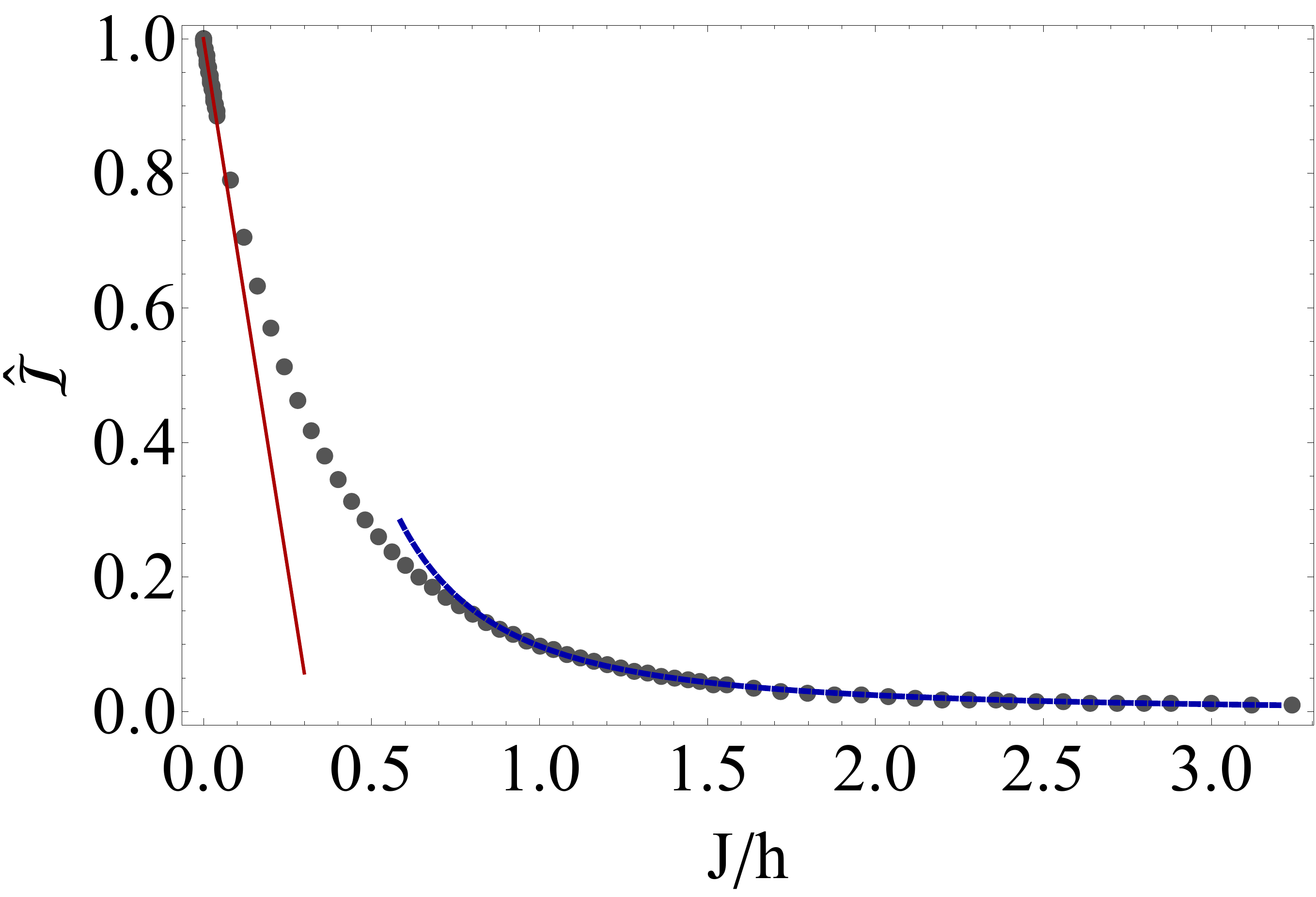}
  \captionof{figure}{\small{Dependence of the remanent density imbalance on the hopping strength $J$ for a chain of non-interacting fermions ($L=100$, $5 \cdot 10^3$ realizations). The continuous red line is the analytical estimate (\ref{RatioSF}) with $J_x = J, J_z=0$. The blue dashed line is a power law fit $a+c (J/h)^{-2}$, with $a=0.003,c=0.101$. %\emph{Inset.} Scaling of the logarithm of the particle imbalance with $\log(J/h)$. The fit is of the form $A \log(J/h)+ B$ with $A=-2.13, B=-2.3$.
   } }
  \label{fig:FitSummary}
\end{figure}

 Fig.~\ref{fig:FitSummary} shows the imbalance as a function of $J/h$, as obtained by exact diagonalization. At small $J/h$ a linear cusp with the slope predicted in~\eqref{RatioSF} (using $J_z = 0, J_x = J$) is seen. For large $J/h$, $\hat{\cal I}$ decays algebraically, as $(J/h)^{-2}$. This scaling can be understood by writing 
\begin{equation}
 \phi_\alpha^2(k) = \frac{x^{\alpha}_k}{ \xi}e^{- \frac{|k-r_\alpha|}{\xi}},
\end{equation}
where $r_\alpha$ denotes the localization center of  $\phi_\alpha$, $\xi$ its localization length (we are neglecting its energy dependence), and the $x^{\alpha}_k$ are positive random variables of $O(1)$ that capture the fluctuations of the squared amplitudes under the exponentially decaying envelope. Partitioning the chain into segments of length $l = \left \lfloor{ \xi}\right \rfloor $ and approximating the $x_k^\alpha$ as uncorrelated variables we obtain \footnote{In~\cite{Bloch2015} a different scaling of the form $1/ \xi^2$ was obtained for the same quantity. The discrepancy with Eq.~\ref{OneOver} arises because the fluctuations of the amplitudes within a correlation length were neglected in that work.}: 
\begin{equation}\label{OneOver}
\begin{split}
 \hat{\mathcal{I}}
 &\approx  \frac{1}{L}\sum_{\alpha=1}^L \tonde{\sum_{R=1}^{L/l}(-1)^{R \, l}\, \frac{e^{- |R-R_\alpha|}}{\xi} \sum_{k=(R-1) l}^{R\, l} (-1)^k x_k^\alpha}^2\\
&\approx \frac{1}{L}\sum_{\alpha=1}^L \tonde{\sum_{R=1}^{L/l}(-1)^{R \,  l}\, \frac{e^{- |R-R_\alpha|}}{\sqrt{\xi}} }^2  \sim \frac{c}{\xi} \sim c\tonde{\frac{J}{h}}^2,
 \end{split}
\end{equation}
where $R_\alpha$ is the block containing the localization center $r_\alpha$, and we have used that in the weak-disorder regime $\xi \sim (J/h)^2\,$\cite{molinari1992}. The scaling~\eqref{OneOver} is verified numerically in Fig. \ref{fig:FitSummary}.

Let us now discuss the qualitative effects of fermionic interactions. 
The addition of a term $ U \sum_{i=1}^L n_i n_{i+1}$ (the equivalent of Ising interactions) to the Hamiltonian~\eqref{Anderson} may have opposite effects, depending on the value of $J/h$. For $J/h \ll 1$, the interaction broadens the distribution of the energy denominators, and thus acts as an additional source of disorder, which reduces the deviation of $\langle \hat{I} \rangle _{\text{dis}}$ from the classical limit. The same holds in the magnetic analogue as confirmed by Eq.~\eqref{RatioSF}. For larger $J/h>1$, the single particle localization length becomes substantial. The dominant effect of interactions is then to mediate (virtual) scattering between single particle states, as discussed in Ref.~\cite{gornyi2005interacting,Basko:2006hh}. One expects that this suppresses the remanent imbalance, as was indeed observed in the experiments of Ref.~\cite{Bloch2015}. For large enough interactions the inelastic scattering processes induce delocalization, as reflected by a breakdown of the locality of the conserved quantities $I_k$. One expects the order parameter to vanish at a $U$-dependent critical hopping $J^*(U)/h$, approaching zero exponentially fast in the inverse of the many-body localization length \cite{Potter2015UniversalProperties, Enss2016InfiniteChains}. The perturbative arguments in~\cite{gornyi2005interacting, Basko:2006hh, rms-IOM} predict that the localized phase is stable for $U< U^*$, where $U^* \propto \delta_\xi/\log \tonde{\mathcal{W}/ \delta_\xi}$ with $\mathcal{W}$ the total bandwidth of the non-interacting Hamiltonian~\eqref{Anderson} and $\delta_\xi$ the average energy gap between single-particle states localized within the same region of size $\xi$, which is assumed to be much larger than the lattice constant, $\xi \gg a$. This corresponds to $J/h \gg 1$, implying $\mathcal{W} \approx J$ and $\xi \approx (J/h)^2$. However, as was pointed out recently \cite{Mirlin2016spectral}, all these studies neglected the phenomenon of spectral diffusion~\cite{Burin2005spectral, Gutman2016EnergyTransport}, which significantly reduces the critical interaction strength in the weak disorder limit to  $U^* \propto \delta_\xi \tonde{\delta_\xi/ \mathcal{W}}^\alpha$ with a positive exponent $\alpha=O(1)$ (up to logarithmic corrections).

\paragraph{Discussion and conclusion.}
We have proposed and analyzed the presumably simplest possible protocol for quantum magnets to exhibit the absence of ergodic dynamics, and thus Many-Body Localization in the form of remanent magnetization in initially ferromagnetically polarized antiferromagnets. 
The present calculation illustrates how the perturbative construction of conserved quantities allows one to make analytic predictions for quantities of experimental relevance. 

Our explicit recipe for constructing the conserved quantities is an analytical alternative to several recent numerical schemes based on DMRG~\cite{YuMPS2015,KennesMPS2015, Khemani2015obtaining, LimMPS2015} or quantum Monte Carlo~\cite{Pollet2016} that allow one to study properties of specific MBL eigenstates. Since the simple formula~\eqref{finalExp} is derived under the sole assumption that the conserved operators have spectrum $\pm 1$, it could be applied to the conserved pseudo-spins constructed numerically in Refs.~\cite{monthus2015integrals, rademaker2015, Pekker2016FixedPoints} for non-perturbative interactions by means of renormalization procedures or diagonalizing flows. 
%However, the approach discussed here allows for analytical insight. 

It would be interesting to extend this calculation beyond the lowest orders, exploiting for example the expansion for the conserved quantities in the forward approximation~\cite{rms-IOM, fpc2016Forward}, to discuss the behavior of the remanent magnetization when approaching the delocalization threshold. An interesting question is whether at the delocalization transition, i.e. at criticality, the order parameter~\eqref{OneOver} exhibits a non-trivial scaling with the system size, potentially reflecting aspects of the expected multifractality of critical wave functions. 
%We leave these questions to future work.

\paragraph{Acknowledgments.} 
Part of this work was done at the Kavli Institute for Theoretical Physics at the University of California Santa Barbara and supported in part by the National Science Foundation under Grant No. NSF PHY11-25915. V. Ros thanks the Paul Scherrer Institute in Villigen for the hospitality and support. 
\bibliographystyle{apsrev4-1}
\bibliography{bibStrFac}

\newpage

\onecolumngrid

\appendix

\section{Appendix A: Derivation of Eq.~(6)}\label{appendixA}

The operator in Eq.~\eqref{def3y} is rewritten as
\begin{equation}
 \sum_{\alpha}  P_\alpha \sigma^z_j P_\alpha= \sum_{i_1= \pm 1} \sum_{i_2=\pm1} \cdots \sum_{i_L=\pm 1}    \prod_{k=1}^L  P(i_k)\;\sigma_j^z \prod_{k=1}^L  P(i_k),
\end{equation}
where we introduced the projectors:
\begin{equation}
 P(i_k)\equiv \frac{1+i_k I_k}{2}.
\end{equation}

To derive Eq.~\eqref{CommutatorsFormula}, we make use of the following operator identities:
\begin{equation}\label{eq:OpId}
\begin{split}
 A B= B A + \quadre{A,B}, \quad   \quadre{A, \prod_{k=1}^L B_k}= \sum_{k_1=1}^L \tonde{ \prod_{k=1}^{k_1-1} B_k} \quadre{A, B_{k_1}}\tonde{ \prod_{k=k_1+1}^L B_k}.
 \end{split}
\end{equation}
For
\begin{equation}
 A^{(1)}= \sigma^z_j, \quad B^{(1)}=\prod_{k=1}^L B_k= \prod_{k=1}^L P(i_k),
\end{equation}
the above identities imply
\begin{equation}
 \prod_{k=1}^L P(i_k) \; \sigma_j^z \prod_{k=1}^L P(i_k)=\prod_{k=1}^L P(i_k) \tonde{\sigma^z_j +\sum_{k_1=1}^L \quadre{\sigma_j^z, \frac{i_{k_1} I_{k_1}}{2}}\prod_{k=k_{1}+1}^{L} P(i_k)}.
\end{equation}
Applying~\eqref{eq:OpId} once more with \begin{equation}                                             A^{(2)}=\quadre{\sigma^z_j, \frac{i_{k_1} I_{k_1}}{2}}, \quad B^{(2)}= \prod_{k=k_{1}+1}^{L} P(i_k)                                                       \end{equation}
yields
\begin{equation}
 \prod_{k=1}^L P(i_k) \;\quadre{\sigma_j^z, \frac{i_{k_1}I_{k_1}}{2}} \prod_{k=k_1+1}^L P(i_k)=\prod_{k=1}^L P(i_k) \tonde{\quadre{\sigma^z_j,\frac{i_{k_1}I_{k_1}}{2}} +\sum_{k_2=k_1+1}^L \quadre{\quadre{\sigma_j^z, \frac{i_{k_1} I_{k_1}}{2}}, \frac{i_{k_2} I_{k_2}}{2}}\prod_{k=k_{2}+1}^{L} P(i_k)}.
\end{equation}
Further iteration with \begin{equation}                                             A^{(n)}=\quadre{\quadre{ \quadre{\sigma^z_j, \frac{i_{k_1} I_{k_1}}{2}},\cdots},\frac{i_{k_{n-1}} I_{k_{n-1}}}{2} }, \quad B^{(n)}= \prod_{k=k_{n-1}+1}^{L}P(i_k)                                                      \end{equation}
finally leads to
\begin{equation}
 \prod_{k=1}^L P(i_k)\; \sigma_j^z \prod_{k=1}^L P(i_k)=\prod_{k=1}^L P(i_k)\tonde{\sigma^z_j +\sum_{N=1}^L\sum_{k_N > \cdots >k_1} \quadre{\quadre{ \quadre{\sigma_j^z, \frac{i_{k_1} I_{k_1}}{2}}, \cdots }, \frac{i_{k_N} I_{k_N}}{2}}}.
\end{equation}
The identity~\eqref{CommutatorsFormula} is established using that $i_k \in \grafe{\pm 1}$ and that
\begin{equation}
 \sum_{i_k= \pm 1} P(i_k)= \mathds{1}.
\end{equation}

\section{Appendix B: Perturbative expressions for conserved quantities}\label{appendixB}
As argued in~\cite{rms-IOM}, the formal expression for the operators in~\eqref{formalPT} reads: 
  \begin{equation}
\begin{split}\label{eq:PertExpOp}
 \delta I^{(n)}_k &= i \lim_{\eta \to 0} \int_0^\infty d\tau e^{-\eta \tau} e^{i H_0 \tau} \quadre{H_1, \delta I_k^{(n-1)}} e^{-i H_0 \tau}+ \Delta I_k^{(n)},
 \end{split}
\end{equation}
where in this case
\begin{equation}
 \begin{split}
  H_0&= \sum_i \tonde{h_i \sigma^z_i - J_z \sigma^z_i \sigma^z_{i+1}},\\
  H_1&= -\sum_i J_x \sigma^x_i \sigma^x_{i+1},
 \end{split}
\end{equation}
and the operator $\Delta I_k^{(n)}$ is a suitable polynomial in the $\sigma^z_i$, such that $I^2_k=\mathds{1}$ is satisfied at the given order in $H_1$. Neglecting the $\Delta I_k^{(n)}$ at any order leads to a modified operator that is still conserved, although it does not have binary spectrum.

 We now discuss how the perturbative expansion needs to be re-summed in presence of resonances. Let $k, k+1$ be the sites giving rise to a first order resonance, i.e., to a small denominator for a particular choice of $\tau, \rho$ in~\eqref{op3}. 
The first-order truncation 
\begin{equation}
 \hat{I}_k=\sigma^z_k + \delta I_k^{(1)}= \sigma^z_k + \sum_{\rho, \tau \pm 1} \tonde{A^{(k)}_{\rho \tau}  O^{(k)}_{\rho \tau}+ B^{(k)}_{\rho \tau} \Delta^{(k)}_{\rho \tau} }
\end{equation}
exactly commutes with the reduced Hamiltonian
\begin{equation}
H^{(k)} = H_0-J_x \sigma^x_k \sigma^x_{k+1}= H_0 - J_x \sum_{\rho, \tau= \pm 1} \tonde{O^{(k)}_{\rho \tau}+ \Delta^{(k)}_{\rho \tau}} \equiv H_0 + H_1^{(k)},
\end{equation}
where $O^{(k)}_{\rho \tau},\Delta^{(k)}_{\rho \tau}$ are given in \eqref{op2}. This can be deduced from \eqref{eq:PertExpOp} setting $H_1 \to H_1^{(k)}$ and $\Delta I_k^{(n)}=0 \; \; \forall \, n$, noticing that 
%the first order correction 
%and $\Delta I_k^{(n)}=0$ for any $n$, by notixing  The first order  commutator
%\begin{equation}
%  \quadre{H_1^{(k)}, \sigma_k^z}= 2 J_x \tonde{\sigma^+_k \sigma^-_{k+1}-\sigma^-_k \sigma^+_{k+1}+\sigma^+_k \sigma^+_{k+1}-\sigma^-_k \sigma^-_{k+1}}
%\end{equation}
%commutes with $H_1^{(k)}$, 
$\quadre{H_1^{(k)}, \delta I_k^{(1)}}=0$ and thus that the perturbative expansion terminates at the first order. To impose the binarity of the spectrum, it is necessary to reintroduce the terms $\Delta I_k^{(n)}$  in order to cancel the terms $\hat{I}^2_k- \mathds{1}$, which are proportional to:
\begin{equation}\label{eq:ToCancel}
 \begin{split}
  \tonde{\sigma_k^+ \sigma_{k+1}^+ + h.c.}^2&=\frac{1+ \sigma^z_k \sigma^z_{k+1}}{2}= P^{(k)}_{1,1}+P^{(k)}_{-1,-1},\\
    \tonde{\sigma_k^+ \sigma_{k+1}^- + h.c.}^2&=\frac{1 -\sigma^z_k \sigma^z_{k+1}}{2}=P^{(k)}_{1,-1}+P^{(k)}_{-1,1},
 \end{split}
\end{equation}
where we defined
\begin{equation}
 P^{(k)}_{\rho, \tau}= \frac{1 + \rho \, \sigma^z_{k}}{2} \frac{1 + \tau \, \sigma^z_{k+1}}{2}.
\end{equation}
The observation that
\begin{equation}\label{eq:ForCancel}
\begin{split}
\frac{1}{2}\grafe{\sigma^z_k, P^{(k)}_{1,-1}-P^{(k)}_{-1,1}}&=\frac{1}{2}\grafe{\sigma^z_k, \frac{\sigma^z_k-\sigma^z_{k+1}}{2}}=P^{(k)}_{1,-1}+P^{(k)}_{-1,1},\\
   \frac{1}{2}  \grafe{\sigma^z_k, P^{(k)}_{1,1}-P^{(k)}_{-1,-1}}&= \frac{1}{2}\grafe{\sigma^z_k, \frac{\sigma^z_k+\sigma^z_{k+1}}{2}}=P^{(k)}_{1,1}+P^{(k)}_{-1,-1},
\end{split}
 \end{equation}

suggests to introduce the modified operator:  
\begin{equation}\label{eq:NewAnsatz}
 \begin{split}
  \tilde{I}_k= \sigma^z_k + &\sum_{\rho \tau= \pm 1} \tonde{\tilde{A}^{(k)}_{\rho \tau} O^{(k)}_{\rho \tau}+C^{(k)}_{\rho \tau} K^{(k)}_{\rho \tau}}+\sum_{\rho \tau= \pm 1} \tonde{\tilde{B}^{(k)}_{\rho \tau} \Delta^{(k)}_{\rho \tau} +D^{(k)}_{\rho \tau} J^{(k)}_{\rho \tau}},
 \end{split}
\end{equation}

with
\begin{equation}
 \begin{split}
   K^{(k)}_{\rho \tau}&= \frac{1 + \rho \, \sigma^z_{k-1}}{2}\quadre{P^{(k)}_{1,-1}-P^{(k)}_{-1,1}}\frac{1 + \tau \, \sigma^z_{k+2}}{2},\\
        J^{(k)}_{\rho \tau}&=\frac{1 + \rho \, \sigma^z_{k-1}}{2}\quadre{P^{(k)}_{1,1}-P^{(k)}_{-1,-1}}\frac{1 + \tau \, \sigma^z_{k+2}}{2}.
    \end{split}
\end{equation}
%and 
%\begin{equation}
% P^{(k)}_{\rho, \tau}= \frac{1 + \rho \, \sigma^z_{k}}{2} \frac{1 + \tau \, \sigma^z_{k+1}}{2}.
%\end{equation}
The condition $\quadre{\tilde{I}_k, H^{(k)}}=0$ imposes the constraints:
\begin{equation}\label{eqUno}
\begin{split}
 \tilde{A}^{(k)}_{\rho \tau} \tonde{h_k -h_{k+1}+ J_z (\tau-\rho)}+  J_x \tonde{1+C^{(k)}_{\rho \tau}}&=0\\
 \tilde{B}^{(k)}_{\rho \tau} \tonde{h_k +h_{k+1}- J_z (\tau+\rho)}+  J_x \tonde{1+D^{(k)}_{\rho \tau}}&=0, 
 \end{split}
\end{equation}
from which Eqs.~\eqref{op3} are recovered for $C^{(k)}_{\rho \tau}=0=D^{(k)}_{\rho \tau}$. This follows from:
\begin{equation}
 \quadre{\tilde{I}_k, H^{(k)}}= \quadre{\sigma^z_k, H_1^{(k)}} +  \sum_{\rho \tau= \pm 1} \tonde{\quadre{C^{(k)}_{\rho \tau} K^{(k)}_{\rho \tau} +D^{(k)}_{\rho \tau} J^{(k)}_{\rho \tau}, H_1^{(k)}}+ \quadre{\tilde{A}^{(k)}_{\rho \tau} O^{(k)}_{\rho \tau}+ \tilde{B}^{(k)}_{\rho \tau} \Delta^{(k)}_{\rho \tau}, H_0}},
\end{equation}
together with:
\begin{equation}
\begin{split}
 \quadre{\sigma_k^+ \sigma_{k+1}^- + \sigma_k^- \sigma_{k+1}^+, H_0}&= -2 \quadre{h_k -h_{k+1} -J_{z}(\sigma^z_{k-1}- \sigma^z_{k+1})} \tonde{\sigma_k^+ \sigma_{k+1}^- - \sigma_k^- \sigma_{k+1}^+},\\
  \quadre{\sigma_k^+ \sigma_{k+1}^+ + \sigma_k^- \sigma_{k+1}^-, H_0}&= -2 \quadre{h_k +h_{k+1} - J_{z}(\sigma^z_{k-1}+ \sigma^z_{k+1})} \tonde{\sigma_k^+ \sigma_{k+1}^+ - \sigma_k^- \sigma_{k+1}^-},\\
 \quadre{\sigma_k^z, H_1^{(k)}}&= -2 J_x \tonde{\sigma_k^+ \sigma_{k+1}^- - \sigma_k^- \sigma_{k+1}^+ +\sigma_k^+ \sigma_{k+1}^+ - \sigma_k^- \sigma_{k+1}^-}, \\
 \quadre{\sigma_{k+1}^z, H_1^{(k)}}&= -2 J_x \tonde{-\sigma_k^+ \sigma_{k+1}^- + \sigma_k^- \sigma_{k+1}^+ +\sigma_k^+ \sigma_{k+1}^+ - \sigma_k^- \sigma_{k+1}^-}.
\end{split}
\end{equation}
Using \eqref{eq:ToCancel} and \eqref{eq:ForCancel}, 
we obtain that $\tilde{I}_k^2=\mathds{1}$ is satisfied provided
\begin{equation}\label{eqDue}
 \begin{split}
  \tonde{\tilde{A}^{(k)}_{\rho \tau}}^2+  \tonde{C^{(k)}_{\rho \tau}}^2 + 2 C^{(k)}_{\rho \tau} &=0,\\
    \tonde{\tilde{B}^{(k)}_{\rho \tau}}^2+  \tonde{D^{(k)}_{\rho \tau}}^2 + 2 D^{(k)}_{\rho \tau} &=0,
 \end{split}
\end{equation}
for each choice of $\tau, \rho = \pm 1$. It can be checked that Eqs.\eqref{eqUno},~\eqref{eqDue} are solved by:
\begin{equation}
 \begin{split}
  \tilde{A}^{(k)}_{\rho \tau}&=-\frac{J_x}{\tonde{\quadre{h_{k}-h_{k+1}+ J_z(\tau-\rho)}^2+ J_x^2}^{1/2} },\\
    C^{(k)}_{\rho \tau}&=-1+\frac{h_{k}-h_{k+1}+ J_z(\tau-\rho)}{\tonde{\quadre{h_{k}-h_{k+1}+ J_z(\tau-\rho)}^2+ J_x^2}^{1/2} },\\
       \tilde{B}^{(k)}_{\rho \tau}&=-\frac{J_x}{\tonde{\quadre{h_{k}+h_{k+1}- J_z( \tau+ \rho)}^2+ J_x^2}^{1/2} },\\
    D^{(k)}_{\rho \tau}&=-1+\frac{h_{k}+h_{k+1}-J_z( \tau+ \rho)}{\tonde{\quadre{h_{k}+h_{k+1}- J_z( \tau+ \rho)}^2+ J_x^2}^{1/2} }.
 \end{split}
\end{equation}

Similar expressions are obtained for the operator $\tilde{I}_{k+1}$.

%\begin{equation}
% \begin{split}
%  \tonde{\sigma_k^+ \sigma_{k+1}^+ + h.c.}^2&=\frac{1}{2}\grafe{\sigma^z_k, \frac{\sigma^z_k +\sigma^z_{k+1}}{2}}= P^{(k)}_{1,1}+P^{(k)}_{-1,-1}= \frac{1+ \sigma^z_k \sigma^z_{k+1}}{2},\\
%    \tonde{\sigma_k^+ \sigma_{k+1}^- + h.c.}^2&=\frac{1}{2}    \grafe{\sigma^z_k, \frac{\sigma^z_k -\sigma^z_{k+1}}{2}} =P^{(k)}_{1,-1}+P^{(k)}_{-1,1}= \frac{1 -\sigma^z_k \sigma^z_{k+1}}{2},
% \end{split}
%\end{equation}

\end{document}